# Ballistic emission spectroscopy and imaging of a buried metal-organic interface

Cedric Troadec,[a]  Linda Kunardi,[a]  and N. Chandrasekhar[a],[b]

[a] Institute of Material Research and Engineering (IMRE), 3 Research Link, Singapore, 117602, Singapore
[b] Department of Physics, National University of Singapore, 2, Science Drive 3, Singapore 117542, Singapore

**Abstract**

The silver-p-phenylene (Ag-PPP) interface is investigated using ballistic electron emission microscopy (BEEM).  Multiple injection barriers and spatial nonuniformity of carrier injection over nanometer length scales are observed. No unique injection barrier is found. Physical reasons for these features are discussed. BEEM current images and the surface topography of the silver film are uncorrelated.

With the growing interest in electronic devices using molecules and polymers as active materials, questions about the properties of interfaces between metals and organics have become critical for understanding the basic physics of their operation [1]. Until now, issues pertaining to the properties of interfaces between metals and organics have been addressed by conventional spectroscopy and current-voltage measurements [2,3]. These techniques average over several millimeters whereas experimental devices are tens of nanometers. We use ballistic electron emission microscopy (BEEM) to investigate a buried metal-organic (MO) interface. BEEM has been used to study metal-inorganic semiconductor interfaces and heterostructures extensively [4,5].

The device configuration for BEEM is well known [4,5]. A semiconductor is overlaid with a thin metal film (typically < 10 nm, termed the base), with an ohmic contact on the opposite side (termed the collector). The top metal film is grounded, and carriers are injected into it using a scanning tunneling microscope (STM) tip. When the energy of the carriers exceeds the Schottky/injection barrier, they propagate into the semiconductor and can be collected from the contact at the bottom. Typically the tunneling current is attenuated by a factor of 1000.

In this work, p-phenylene (PPP), a blue emitter with high photoluminescence efficiency [6], is used as the semiconductor. The HOMO and LUMO levels are 5.2 and 2.1 eV respectively. PPP can be evaporated, with the films having small conjugation lengths of between 4 and 7 phenyl rings for low substrate temperatures [7]. The choice of the base depends on the type of the injection barrier (whether electron or hole) and the alignment of levels to be measured. In this work we have chosen Ag since it has been shown to yield injection-limited contacts for hole injection into the polyparaphenylene/vinylene family of organics [8]. Noble metals do not damage organics, and diffusion of the metal into the organic is minimized when deposited onto a sample held at low temperature.

A low temperature home-assembled STM system and a separate sample preparation chamber were used to carry out the experiments. The samples consist of a pre-deposited gold film on glass or silicon, which serves as the collector. A nominally 100 nm thick PPP film is evaporated at a pressure below $5 \times 10^{-7}$ mbar, onto the substrate cooled to 77 K. After removal from vacuum, a mechanical mask is used to define a diode area of ~ 2 $mm^2$. A nominally 10 nm thick Ag film is then evaporated under identical conditions to serve as the base. The sample is then placed in the STM chamber, where it is cooled to 77 K for the experiments. The current noise of the setup is typically 1 pA.

Ex-situ SEM (Scanning Electron Microscope) and TEM (Transmission Electron Microscope) studies of the PPP films were done. The SEM images (not shown) are unremarkable and indicate a reasonably flat film, with no pinholes or protrusions. The TEM images indicate the presence of domains approximately 50 nm wide and several hundred nanometers long, as shown in Fig. 1(a). Diffraction confirms an amorphous film. When comparing the energy level alignment, as shown in Fig. 1(b), the Schottky-Mott rule yields a barrier of 0.9 V, with a work function of 4.3 eV for Ag. The real situation, due to the image force correction, is as shown by the dotted line, i.e. the barrier is lowered by 15 meV, and shifted inside the organic by 300 Å.

Over two hundred individual I-V's, acquired over different locations within a nominally 25 nm square, are averaged in order to improve the signal to noise ratio. Repeated acquisition of spectra at the same point was found to be detrimental to the sample, as evidenced by the increased instability of the spectrum with time. The Schottky barrier is the point where the collector current is non-zero. Thermal effects can cause significant deviations from the true value, but are negligible at 77 K. It is important to ensure that the Ag film is reasonably flat. Unless this requirement is met, attempts to tunnel into patches of the metal film, which are poorly connected, can lead to tip crashes. Lateral variations in the Fermi level due to poor connectivity would give incorrect Schottky barriers [4,5]. Early BEEM experiments on MO interfaces yielded low currents [9]. Spin-coated samples consistently yield low currents, barely above the noise.

Fig. 2 shows a BHES, after averaging, for the PPP/Ag interface, with a typical raw spectrum as the inset. The current-voltage (I-V) spectroscopy is taken over a range of 0 to 2 V. Qualitatively, this curve is similar to the BEEM spectra seen for metal-inorganic semiconductor (MIS) interfaces. The injection barrier may be taken to be the value of the voltage where the collector current begins to deviate from zero. The raw spectrum shows that this deviation occurs near 1 eV, a value close to the Schottky-Mott model. However, this is deceptive, as we will show below. A more sophisticated technique to obtain the Schottky barrier is to obtain a dI/dV, also in Fig. 2, so that the steps corresponding to the injection barrier can be readily discerned. It is not clear whether this technique can be applied to MO interfaces; hence the derivative should be interpreted with caution. The dI/dV exhibits multiple steps at voltages of 0.6 V, 0.9 V and 1.1 V. This indicates that the injection barrier is not a unique quantity, in contrast to the case of Au-Si interfaces. We mention that multiple steps in the dI/dV have been observed for MIS junctions as well (eg. GaAs and GaN) [4,5].

Extraction of the Schottky barrier height from the raw BHES data is done using the Bell and Kaiser [10] (BK) model, which uses a planar tunneling formalism. Transverse momentum (k vector) conservation is clearly inapplicable at the MO interface. The functional dependence of the BEEM current on voltage is a power law $(V-V_0)^n$, where $V_0$ is the injection barrier. The best fit to our data is obtained with an exponent of 7/2 and a $V_0$ ranging from 0.3 to 0.5 V. This value is clearly much smaller than the barrier deduced from the Schottky-Mott model. Conformation changes in PPP, known to occur at T > 20K, can partly explain this deviation, since torsional oscillations of the phenyl rings cause shifts in the HOMO and LUMO and also change the bandgap [11]. This should be contrasted with the barrier heights inferred from the dI/dV, which are 0.6, 0.9 and 1.1 eV.

An STM image of the top Ag film, at 0.5 V and 1 nA is shown in Fig. 3(a). The topography scale is 1.5 nm showing that the Ag film is well connected. The BHES I-V and its derivative enable a choice of suitable imaging conditions. For instance, a bias voltage of 0.8 V should yield measurable collector currents. An image obtained at 0.8 V (0.2 V above the first dI/dV threshold of 0.6 V) is shown in Fig. 3(b). Both images are 50 nm square. We find that the BEEM current image does not correlate with the STM derivative (not shown), which is readily apparent from the images. The image indicates non-uniform transparency of the interface. Bright spots indicate regions that are more transparent. The size of such regions appears is a few nanometers. In MIS interfaces, the STM derivative and the BEEM images often correlate well, since the BEEM current increases in the vicinity of grain boundaries where the density of surface states is small. For MO interfaces, poor correlation of the BEEM and STM derivative images suggests that surface states are insignificant.

In conclusion, we demonstrate how BEEM can be used for high resolution studies of buried metal-organic interfaces. Considerable deviation of the barrier magnitude from the Schottky-Mott rule is observed. Images of the interface transparency show substantial lateral/spatial non-uniformity. Naturally, this kind of investigation needs to be extended to other organics and metals in order to obtain a better understanding of charge transfer across MO interfaces.

## ACKNOWLEDGMENTS

NC thanks Drs. V. Narayanmurti, V. I. Arkhipov, C. Joachim, A. Dodabalapur and I. Shalish for discussions. This work was supported by A*STAR and IMRE, Singapore.

List of Figures

Figure 1a. TEM image of a nominally 100 nm thick PPP film.

Figure 1b. Schottky barrier for the Ag-PPP interface using alignment of vacuum levels, i.e. Schottky-Mott rule.

Figure 2. The I-V and the dI/dV of the Ag-PPP interface. Multiple thresholds evidenced as steps in the dI/dV are clearly seen. Inset shows one single spectrum as recorded.

Figure 3. STM and BEEM current images (a) STM image of a 50 nm square region. The topography scale is 1.2 nm (b) BEEM current image of the corresponding region at 0.8 V bias, with a full scale of 3.5 pA.

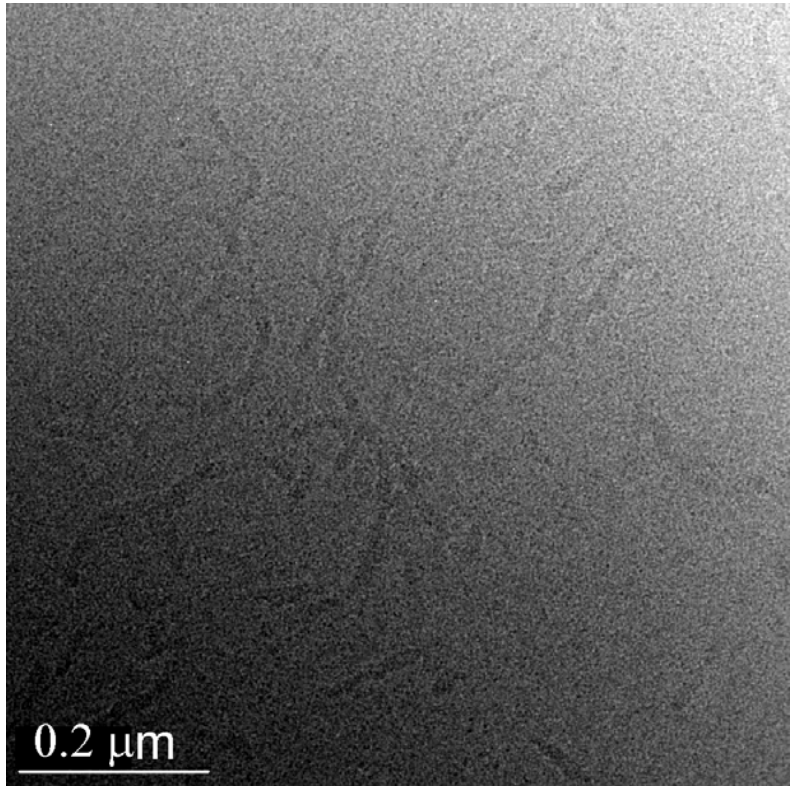

**Figure 1(a).**

**Figure 1(b).**

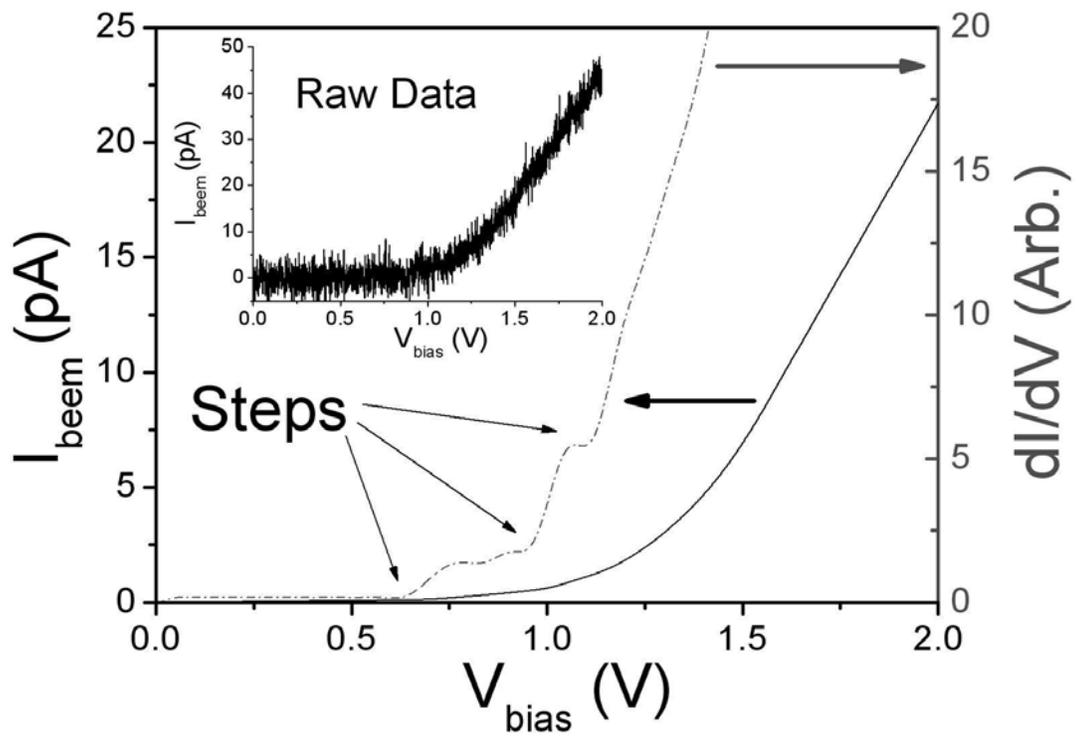

**Figure 2.**

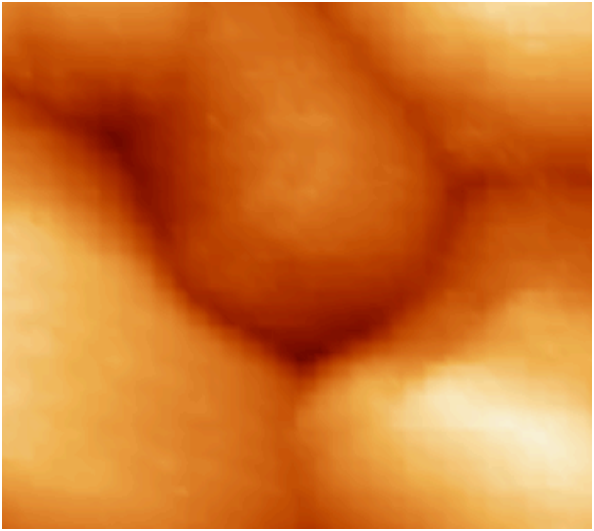 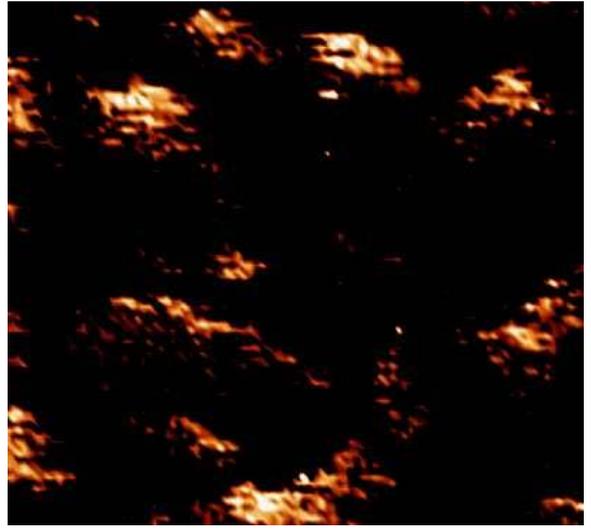

**Figure 3(a)**                                            **Figure 3(b)**